\documentclass[aps,twocolumn,prx,floatfix,superscriptaddress,showpacs,longbilbiography]{revtex4-1}
\usepackage[utf8]{inputenc}

\usepackage{graphicx}
\usepackage{dcolumn}
\usepackage{bm}

\usepackage{braket}
\usepackage{amsfonts}
\usepackage{amsmath}
\usepackage{amssymb}

\usepackage{color,soul}
\usepackage{wasysym}
\usepackage{mathrsfs}
\usepackage{times}

\usepackage[dvipsnames,svgnames,table]{xcolor}
\usepackage{hyperref}
\usepackage{fancyhdr}
\usepackage{float}
\usepackage[english]{babel}
\usepackage[caption=false]{subfig}
\usepackage{braket}

\hypersetup{
pdfstartview={FitH},
colorlinks=true,    
linkcolor=NavyBlue, 
citecolor=Maroon,   
filecolor=NavyBlue, 
urlcolor=NavyBlue   
}

\newcommand{\bea}{\begin{eqnarray}}
\newcommand{\eea}{\end{eqnarray}}

\begin{document}
\title{Wavefunction collapse driven by non-Hermitian disturbance}

\author{Jorge Martínez Romeral}
\affiliation{Catalan Institute of Nanoscience and Nanotechnology (ICN2), CSIC and BIST, Campus UAB, Bellaterra, 08193 Barcelona, Spain}
\affiliation{Department of Physics, Universitat Autònoma de Barcelona (UAB), Campus UAB, Bellaterra, 08193 Barcelona, Spain}
\author{Luis E. F. Foa Torres}
\affiliation{Departamento de F\'{\i}sica, Facultad de Ciencias F\'{\i}sicas y Matem\'aticas, Universidad de Chile, Santiago, 837.0415, Chile}
\author{Stephan Roche}
\affiliation{Catalan Institute of Nanoscience and Nanotechnology (ICN2), CSIC and BIST, Campus UAB, Bellaterra, 08193 Barcelona, Spain}
\affiliation{ICREA--Instituci\'o Catalana de Recerca i Estudis Avan\c{c}ats, 08010 Barcelona, Spain}

\begin{abstract}
In the context of the measurement problem, we propose to model the interaction between a quantum particle and an “apparatus” through a non-Hermitian Hamiltonian term. We simulate the time evolution of a normalized quantum state split into two spin components (via a Stern-Gerlach experiment) and that undergoes a wavefunction collapse driven by a non-Hermitian Hatano-Nelson Hamiltonian. We further analyze how the strength and other parameters of the non-Hermitian perturbation influence the time-to-collapse of the wave function obtained under a Sch\"{o}dinger-type evolution.
We finally discuss a thought experiment where manipulation of the apparatus could challenge standard quantum mechanics predictions.
\end{abstract}

\date{\today}
\maketitle

\textbf{Introduction}. A long-standing mystery of quantum mechanics foundations is related to the ultimate nature of the state vector and the mechanism driving the transition of the particle state from the quantum to classical world, materialized by the measurement event, and giving rise to the “measurement problem”.  It is a cornerstone puzzle which roots all debates and multiple views concerning the interpretation of quantum mechanics and the ultimate access to some unfathomable reality~\cite{EPR1935,Bohr1935,Bell1964,WIGNER198463,Laloe2001,Bell_Aspect_2004,bell_against_1990,RevModPhys.65.803,Hance_2022,ormrod2023theories}.

For a closed system, the time-dependent Schrödinger equation describes the deterministic evolution of the state vector, which by the superposition principle, can be driven by multiple components (and their interference) expressed in a given orthogonal basis set. 
As long as the evolution remains unitary, the state vector maintains its quantum coherence and in general will be a superposition state.
The access to some of the state vector properties (position, momentum, spin, etc) however requires a true measurement, which is introduced through a different dynamical process requiring the collapse of the state vector to a probabilistically selected outcome.
This is the collapse (“projection”) postulate of quantum mechanics and, introduces an irreversible process where information is lost, thereby adding to the cut with the unitary Sch\"odinger evolution in absence of a measurement. Some fundamental questions arise such as (i) \textit{how to properly model the effect of the interaction of the system with an external environment during the measurement process?} (ii) \textit{can we theoretically access the dynamical transient regime between the deterministic evolution of the quantum state vector towards the final state (after measurement/projection) which appears as a reduction of the initial wave-packet?} (iii) \textit{can the measured final outcomes be in any sense dependent on the interaction with the environment/apparatus' properties?}

In absence of a formalism that can follow the time-evolution of any quantum states during the measurement process, those questions have so far remained debated but unsettled. The search for an underlying physical reality of the wavefunction collapse and its relation with the quantum measurement problem belongs to the quest for a deeper understanding of the foundations of quantum mechanics. The main proposals include the Ghirardi-Rimini-Weber (GRW) model \cite{ghirardi_unified_1986,PhysRevA.42.78,BASSI2003257} and the continuous spontaneous localization \cite{pearle_combining_1989}, that establish the knowledgeable frontier on this issue  \cite{laloe_quantum_2019}~\footnote{Related to the GRW scheme, we also mention the so-called quantum-drift approach, which uses a stochastic Schrödinger equation and which is inspired by earlier works in the field of quantum transport by Pastawski and collaborators, see ~\cite{fernandez-alcazar_decoherent_2015} and references therein.}. These two models are mainly aimed at taking out the measurement as something that is being carried out solely by a macroscopic measurement apparatus, but: (i) they add different (stochastic) ingredients to the Schr\"{o}dinger dynamics, (ii) the collapse itself is assumed to happen as an automatic phenomenon driven by ad-hoc dynamical equations, while there is no insight on the underlying physical mechanism that would induce the collapse itself during the time-evolution of the quantum state. Differently, the so-called Diósi-Penrose collapse framework proposes that gravity acts as a driving force to the wavepacket reduction \cite{Diosi1989,Penrose1996,RevModPhys.85.471,Laloe2020,Laloe2022}, a scenario that has been challenged by experiments searching for its radiative signature \cite{Donadi2021,Arnquist2022,Donadi2023}.

Here, we put forward how the collapse of the wavefunction can emerge by introducing a non-Hermitian perturbation to an otherwise Hermitian Hamiltonian. In this model, the quantum dynamics is fully governed by the Schr\"{o}dinger equation. We illustrate this idea using the Hatano model \cite{PhysRevB.56.8651,PhysRevB.58.8384,PhysRevLett.77.570,PhysRevB.108.214308} 
as a non-Hermitian perturbation, we follow numerically how the wavefunction collapses into one of its initial up or down spin state, an event captured in space and time. The initial wavefunction is built from entangled spin-orbital components, as generated typically in a Stern-Gerlach experiment, while the local disturbance introduced by the non-Hermitian term breaks nonlocality and force the wavefunction to be reduced into one of the two possible outcomes. The time-to-collapse is driven by the asymmetry and strength of non-Hermitian Hamiltonian parts and related to an energy scale dictated by the local interaction between the ``apparatus" and the incoming quantum particle trajectories. The collapse mechanism is thus a natural consequence of a local interaction which aids the wavepacket reduction under a Schr\"{o}dinger-type evolution. Although there are still mysteries under the carpet, this path may provide insights which could be experimentally tested, besides the natural access to a physical scale usually missing in the standard formulation of a quantum measurement: the time-to-collapse.

\begin{figure}[tbh]
\includegraphics[width=\columnwidth]{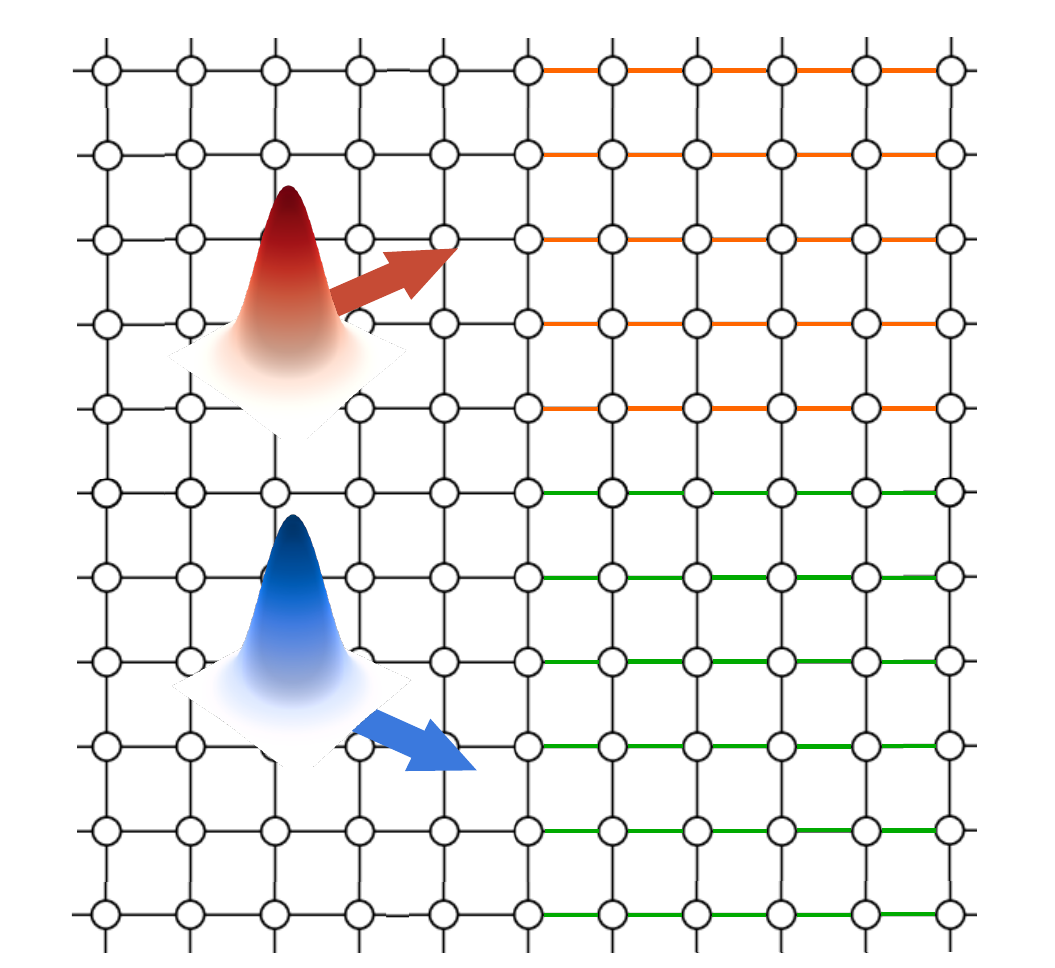}
\caption{\label{fig:scheme} Scheme of the initial state, equation \ref{eq:initialstate}, jointly with the real space structure. The orange and red highlighted regions correspond to the non-Hermitian Hamiltonian with $g_1$ and $g_2$ respectively. The non-coloured region corresponds to a square lattice tight-binding Hamiltonian.}
\label{Fig1}
\end{figure}

\textbf{Non Hermitian Hamiltonian}. %
As a starting point, we choose a square lattice with nearest neighbours hopping term. It is Hermitian $\hat{\cal H}_{0}=\sum_i\gamma_0\left(c_{j}^{\dagger}c_{j+1}+c_{j+1}^{\dagger}c_{j}\right)$ in most part of the real space except in some defined region (see Fig.\ref{Fig1}) where a breaking of hermiticity is introduced through changing hopping amplitude according to the Hatano-Nelson model \cite{PhysRevB.56.8651,PhysRevB.58.8384,PhysRevLett.77.570,Orito2022,PhysRevB.108.214308,Takane2023}, which become asymmetric (non-reciprocal), and whose asymmetry strength is piloted by a parameter g, as follows:

\begin{equation}
    \hat{\cal H}_{nH}=\sum_i\gamma_0\left(e^{-g}c_{j}^{\dagger}c_{j+1}+e^{g}c_{j+1}^{\dagger}c_{j}\right)
\end{equation}
and which creates a real-space asymmetric hopping, modulated by $g$. This model is here limited to an alteration of hopping amplitude along one direction, but the lattice model is two-dimensional. We design the special arrangement as depicted in Fig \ref{fig:scheme}, where two parts of the lattice (mimicking the interaction zone between incoming quantum and the ``measurement apparatus") has different non-Hermitian magnitude, $g_1$ for the upper part and $g_2$ for the lower. In addition, a ``localized'' non-Hermitian form of gains is also studied, defining
\begin{equation}
    \hat{\cal H}_{nH}= \sum_jz\,\phi(\mathbf{x}_j,\mathbf{x}_0)\hat{n}_j,
\end{equation}
where $\phi(\mathbf{x}_j,\mathbf{x}_0)$ are exponential localized weights in $\mathbf{x}_0$, and the $z$ is a pure imaginary coefficient that indicates the strength of the gains in the centre of the Gaussian function.

At initial time, a quantum state is prepared with a Gaussian shape in real space and is an equal superposition of up and down spin components (as for instance  formed during a Stern-Gerlach experiment). The wavepacket starts in the space region where the Hamiltonian is Hermitian, with each spin part having different momenta, one propagating to $x>0$ and $y>0$ while the other propagates to $x>0$ and $y<0$. This reproduces the splitting effect of a Stern-Gerlach apparatus, where an inhomogeneous magnetic field entangles orbital and spin degrees of freedom and splits apart the two components. The total wavefunction at initial time $\ket{\Psi(0)}$ reads: 
\begin{equation}\label{eq:initialstate}
        \ket{\Psi(\mathbf{r}_1,\mathbf{r}_2;\mathbf{k}_1,\mathbf{k}_2}=\dfrac{1}{\sqrt{2}}\left[\psi(\mathbf{r}_1,\mathbf{k}_1)\otimes\ket{\uparrow}+\psi(\mathbf{r}_2,\mathbf{k}_2)\otimes\ket{\downarrow}\right].
\end{equation}
where spatial components are defined as:
\begin{equation}\label{eq:WP}
    \psi(\mathbf{r})=\dfrac{1}{\sqrt{2\pi}\sigma_r}\exp{\left(\dfrac{-|\mathbf{r}-\mathbf{r}_0|}{4\sigma_r^2}\right)}\exp{\left(-i\mathbf{k}_0\cdot\mathbf{r}\right)},
\end{equation}
    defining $\sigma_r$ as the uncertainty in position and $\mathbf{r}_0$ is the initial position where the wave packet is centred. The state is then evolved under the action of the evolution operator ($\ket{\Psi(t)}=e^{-i(\hat{\cal H}_{0}+\hat{\cal H}_{nH})t/\hbar}\ket{\Psi(0)}$), depending on the total Hamiltonian (composed by the Hermitian and the non-Hermitian part). At each time step, after applying the non-Hermitian Hamiltonian the wave packet is renormalized. We then compute the average value of the out-of-plane spin polarization as $\left<S_z(t)\right>=\bra{\Psi(t)}\hat{S}_z\ket{\Psi(t)}$ which is zero at initial time.

\begin{figure*}
\includegraphics[width=0.6\textwidth]{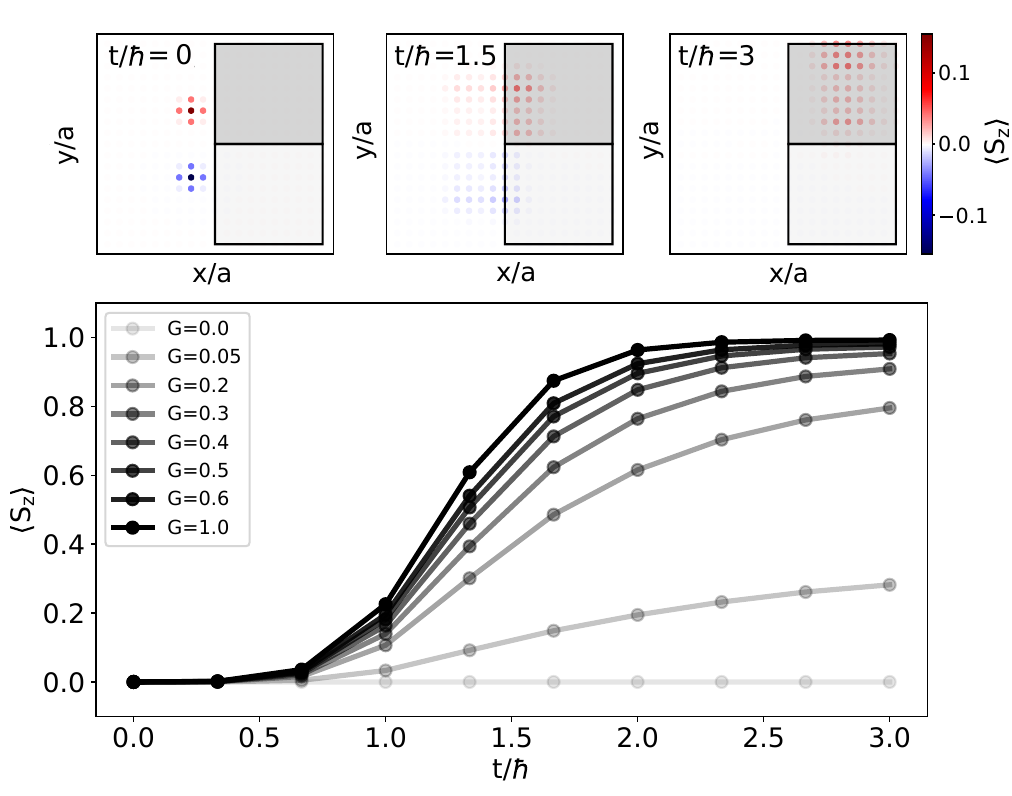}
\caption{Top figures: in blue/red the out-of-plane spin polarization projected in real space, $\braket{\hat{S}_z(\mathbf{r})}$. The two shaded regions are the two non-Hermitian parts with $g_1$ and $g_2$ In this case they correspond to $G=0.5$. From left to right they correspond to different times in the simulation, $t/\hbar$: $0$, $1.5$ and $3$.  Bottom figure: mean value of the out-of-plane spin polarization, $\braket{\hat{S}_z} $ as a function of $t/\hbar$ for different values of the difference in the non-Hermitian magnitude of the top and bottom regions, $G=g_1-g_2$.}
\label{fig:Figure2} 
\end{figure*}

\textbf{Wavefunction collapse}. The dynamics of wavefunction collapse for varying degree of non-Hermitian disturbance are shown in Fig \ref{fig:Figure2}. We simulate the collapse dynamics (equivalently, the increase of $\left<S_z(t)\right>$) as a function of the difference between the non-Hermitian perturbations acting on the two different spatial locations (see Fig.\ref{Fig1}), quantified by $G=g_1-g_2$. It clearly appears that if the initial spin polarization $\left<S_z(t=0)\right>$ is zero (with equal weight of upspin and downspin components), as time evolves, a nonzero $\left<S_z(t)\right>$ emerges (main panel) as the wavefunction is spatially entering into the interaction zone (upper panel), with a progressive collapse of the superposition to one of the outcomes (either upspin or downspin polarization). The collapse is fully realized once $\left<S_z(t)\right>=1$.

Due to the presence of complex eigenvalues with negative imaginary part introduced by the non-Hermitian perturbation, the respective eigenvectors will be amplified in each time step leading to an amplification of the part of the state that is entering the non-Hermitian part with higher $g$. This effectively shows a collapse in both the spin and spatial components. The time of collapse is tunable by changing the difference between both regions as shown in Fig.\ref{fig:Figure2} when decreasing $G$. When $G=0$ both parts have equal non-Hermitian strengths and therefore the spin component does not collapse to any region, although the spatial distribution is altered.

\begin{figure}[tbh]
\includegraphics[width=0.9\columnwidth]{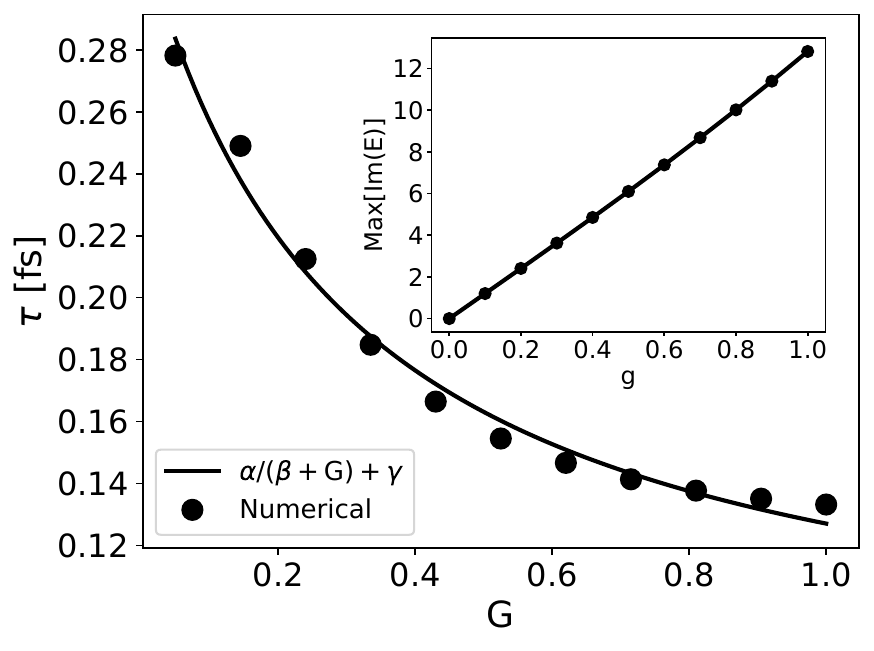}
\caption{Main frame: Collapse time in femtoseconds as a function of the non-Hermitian strength jointly with the inverse function fit
$h(G)=\dfrac{\alpha}{\beta+G}+\gamma$. In this case $\alpha=0.07$ fs, $\beta=0.35$ and $\gamma=0.07$ fs. Inset: Maximum value of the imaginary part of the 2D Hatano-Nelson model as a function of the non-Hermitian strength $g$.}
\label{fig:fit}
\end{figure}

Further ahead, we can fit the behavior of the polarization dynamics showed in Fig.\ref{fig:Figure2} to a logistic function of the form,
\begin{equation}\label{eq:logistic}
    f(t)=\dfrac{a}{1+e^{-b(t+t_0)/\hbar}} +c,
\end{equation}
where $a$ and $c$ are dimensionless and $b$ has units of energy. We can define the collapse time as $\tau=\hbar/b$. We expect the behavior of the collapse to follow this function as it is mainly driven by the amplification of the different eigenstates, with eigenvalues $a,b,c,..$, due to the non-Hermitian time evolution plus the required normalization in each time step. As the out-of-plane spin polarization adds a minus sign to the spin-down part, this leads to the division of the subtraction of exponentials by the sum of exponentials, $\left<S_{\mathrm{z}}\right>=(e^{-a_{\uparrow}t}-e^{a_{\downarrow}t}+e^{-b_{\uparrow}t}-e^{b_{\downarrow}t}+...)/\left(e^{-a_{\uparrow}t}+e^{a_{\downarrow}t}+e^{-b_{\uparrow}t}+e^{b_{\downarrow}t}+...\right)$. In the end the largest eigenvalue, say $\epsilon$, will dominate over the rest gathered in coefficients $C_1$ and $C_2$, so leading to a dominant term of logistic type form, plus other subdominant terms,  $\left<S_{\mathrm{z}}\right>=\left( e^{\epsilon t}+C_1\right)/\left(e^{\epsilon t}+C_2\right)= 1/\left(1+e^{-\epsilon t}C_1\right) +...$
This collapse time is plotted as a function of $G$, jointly with its exponential fit in Fig. \ref{fig:fit} (main frame). As expected, the collapse time decreases as 
$G$ increases and increases when $G$ approaches zero. This behaviour is expected as $G$ measures the non-Hermitian perturbation acting on the incoming quantum states. One 
can relate this collapse time to an energy scale which should be related to the energy exchange during the ``measurement process" and which will depend on the microscopic energy fluctuation profile
dictating the apparatus interaction with the quantum state. Microscopically, the scaling of $\tau$ with G is directly linked to the increase of the imaginary part of the eigenvalues of $\hat{\cal H}_{nH}$ (Fig. \ref{fig:fit} (inset)), of which magnitude dictates the value of $\tau$ during the dynamics. Given that the wavefunction is normalized during the time evolution, the quantum particle will be subjected to this effective energy exchange with the external source at the origin of the non-Hermitian part, breaking unitarity. 

\begin{figure*}
\includegraphics[width=0.9\textwidth]{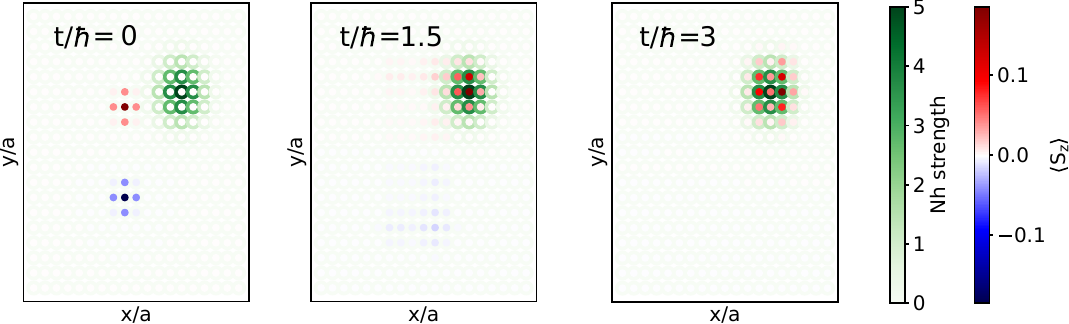}
\caption{\label{fig:Figure3} 
In red/blue is shown the out-of-plane spin polarization projected in real space, $\braket{\hat{S}_z(\mathbf{r})}$ for different times of the simulation, $t/\hbar$: $0$, $1.5$ and $3$. In green it is shown the strength of a localized non-Hermitian part of $z=5\cdot i$.}
\end{figure*}

Finally Fig.\ref{fig:Figure3} shows the case of a spatially localized non-Hermitian part in the upper region of the system. In contrast with Fig. \ref{fig:Figure2} the effective collapse of the spin part is the same while the spatial part varies due to the different spatial distribution of the non-Hermitian Hamiltonian. This situation could represent the interaction between an incoming trajectory of a spin-split state with a local STM tip. As soon as the interaction takes place, the collapse of the wavefunction occurs in a timescale which is determined by the model parameters. In both Figures \ref{fig:Figure2} and \ref{fig:Figure3}, we have taken into account both $g_1$ and $g_2$ equal or greater than zero. Therefore here we only present an amplification effect. Allowing the coefficients to be negative, leads to a loose mechanism in the opposite region. If only one region is negative, the spin polarization won't change enough to lead to an effective collapse. On the other hand, combining negative values with positive values will lead to the effective collapse in the region with larger positive non-Hermitian strength. These simulations illustrate the effect of a spatially localized source of non-Hermitian disturbance on the time evolution of a multi-components wavefunction. It appears clearly that the time-to-collapse is driven by the relative inhomogeneous perturbation in real space which enforces the outcome of the wavefunction at the location where maximum disturbance is created.

\textbf{Discussion and final remarks}. We have proposed the use of a non-Hermitian perturbation as a means of modelling a quantum measurement and the subsequent state vector collapse ~\footnote{Note that after we finalize this study, we became aware of recent works addressing the measurement problem along the same line \cite{Singh_2024,singh2023emulating}}. The perturbation is meant as a poor-man's model of the interaction with the measuring apparatus and allows for a description under a single dynamic equation, the Schr\"odinger equation. To illustrate this, we used a simple model for a Stern-Gerlach type of experiment where the meter can influence either both spatially separated spin parts or a single one. In this particular case, the mechanism for collapse is simple amplification of the selected state, which depends on the parameters of the perturbation Hamiltonian. 

The exploits of such scheme for a quantum measurement are two-fold: 
\begin{itemize}
    \item the split in the description of the dynamics (Schr\"odinger dynamics/irreversible collapse) is substituted by a change in the Hamiltonian when it is perturbed by the measurement (whatever its origin) under a single dynamical equation;
    \item the measurement process acquires a different status as one has to incorporate details or parameters describing the interaction associated to the measurement.
    \end{itemize}\
As a byproduct, there is a natural physical scale that enters the game, the time-to-collapse. In this simple setup shifting the amplification from one part of the beam or the other changes the final outcome. The frequency of each result should satisfy Born's rule if all the predictions of quantum mechanics are to be kept. This seems to indicate that the initial state (just before the measurement) must influence the apparatus. But, can we eventually tune the measurement parameters/apparatus to challenge the usual rules? This would be akin to start playing with loaded dice or even to stop playing dice at all. The existence of such “hidden physical reality” driving the collapse could be tested by adjusting the interaction degree between the two incoming trajectories so as to start to see deviations from Born's rule or even obtain one result with certainty. We could imagine placing some identical trapped ultracold atoms (or qubits) in their ground states or tuned to some excited states to weakly differentiate the energy environment placed onto the multiple quantum particle trajectories, so providing a different source of energy to exchange with (see the discussion in the context of continuous spontaneous localization \cite{Laloe2014}). This type of ``thought experiment" could be materialized thanks to recent developments in Stern-Gerlach interferometry \cite{Machluf2013,Margalit2021,Keil2021,PhysRevLett.130.113602}.

The collapse scheme described here could equally apply to two entangled particles split apart and whose spin-spin correlations would be measured independently, satisfying the violation of the Bell´s inequality \cite{Bell1964,Bell_Aspect_2004,Aspect1982,Aspect2}. The same local physical mechanism (through non-Hermitian perturbation) could just enforce the projection/collapse of the two-particle wavefunction.

Finally, we have to point out that this is just a starting step in the use of non-Hermitian Hamiltonians for the dynamical modelling of quantum measurements as there are many issues that are worth of further investigation. The mechanism described here seems to be the simplest possible one, but it is probably not the one bringing more insights onto the process and the nature of Heisenberg's cut. One could, for example, envision mechanisms using more actively truly non-Hermitian properties such as defectiveness~\cite{heiss_physics_2012,rotter_review_2015,FoaTorres_2020}, the lack of a full set of eigenstates. We therefore see a clear need for more work along this line which may bring a revived interest into this key issue at the foundations of quantum physics.

\begin{acknowledgments}
 ICN2 is funded by the CERCA programme / Generalitat de Catalunya, and is supported by the Severo Ochoa Centres of Excellence programme, Grant CEX2021-001214-S, funded by MCIN / AEI / 10.13039.501100011033. This work is also supported by MICIN with European funds‐NextGenerationEU (PRTR‐C17.I1) and by Generalitat de Catalunya. We thank the Catalan Quantum Academy for support. L.E.F.F.T. acknowledges financial support by FONDECYT (Chile) through grant 1211038 and of The Abdus Salam International Center for Theoretical Physics and the Simons Foundation.
\end{acknowledgments}

\end{document}